# Software developers, moods, emotions, and performance

Studies show that software developers' happiness pays off when it comes to productivity[1]

Daniel Graziotin, Xiaofeng Wang, Pekka Abrahamsson

Free University of Bozen-Bolzano, Italy

High tech companies like Google, Facebook, Supercell and several Silicon Valley startups are well known for their *perks*, such as having fun things to do, good food to eat at working places during working hours. One underlying assumption is "happy software developers perform better than unhappy ones." However, has there ever been any evidence supporting this? Evidence is especially desirable for those who do not work in these "fun" companies, since the paradisiac environment usually described by tech giant workers turns into horror stories told by frustrated developers on a daily basis. We often read such stories on websites like thedailywtf.com and clientsfromhell.net. It seems that managers need to be reminded of the "obvious" claim that happy developers work better.

In scientific articles, it has long been claimed that the best way to improve software developers' performance is by focusing on people [1], and that *people trump process* [2]. Software development is deeply dominated by human factors [3]. Software development is complex and purely intellectual. It is accomplished through cognitive processing abilities [4]. No matter how much we would like to *believe* that software construction processes can be engineered, software developers are creative human beings and they need to make sense of unpredictable, turbulent environments [5]. Sense-making processes are influenced by the *affects* (emotions and moods) of individuals [6], so are several cognitive processes tied to problem-solving performance [7]. In the software engineering research arena very few studies have been conducted on how developers *feel* and what the consequences are of them *feeling*.

This VoE article summarizes the existing psychology and software engineering studies conducted on how software developers experience affects and demonstrates how practitioners are calling for such studies to be conducted.

---

# What scientists know about the affects of software developers

*Affects* is a highly polymorphic term, which means it has multiple meanings. There is not even an agreement in psychology research on what differentiates emotions and moods. While the tendency is to consider moods as prolonged feelings where a stimulus is not immediately identifiable by the subject (whereas emotions have a clear origin), affects can be safely considered as an umbrella term for emotions and moods. Affects are represented using discrete variables (e.g., joy, sadness, anger) or dimensions (e.g., valence, arousal). More on how affects can be represented and measured can be found in [8], [9].

Shaw [10] was perhaps the first to observe that the role of emotions in the workplace has been the subject of management research, but information systems research has limited its focus on job outcomes such as stress, turnover, burnout and satisfaction. By observing 12 senior-level undergraduate students, who were engaged in a semester-long implementation project, he showed that the affects of a software developer may dramatically change during a period of 48 hours.

Lesiuk [11] recruited 56 software engineers to understand the effects of music listening on software design performance. Data was collected over a five-week period. The participants self-assessed affects and design performance. The results indicated that positive affects and self-assessed performance were lowest with no music, while time-on-task was longest when music was removed. Narrative responses revealed the value of music listening for positive mood change and enhanced perception on design while working.

Along a similar line, Khan et al. [4] conducted two studies to understand the impact of affects on the debugging performance of developers. In the first study, positive affects (valence and arousal) were induced to the software developers. Subsequently, they completed a quiz on software debugging. In the second study, the participants wrote traces of the execution of algorithms on paper. During the task, the affect *arousal* was induced in the participants. Overall, the two studies provided empirical evidence for a positive correlation between the affects of software developers and their debugging performance.

Colomo-Palacios et al. [12] aimed to integrate the developers' and the system users' emotions into the process of requirements engineering. They conducted a study on two software projects and 11 individuals. In total, 65 user requirements were produced between the two projects, which lasted six months and seven months, respectively. Each requirement faced tens of revision. Each participant rated the affects associated to each requirement version. The results showed that the affects related to pleasantness associated to the final requirements are higher than for non-final requirements, while the affects related to mental activation (arousal) for the final requirements are lower than for non-final requirements.

Wrobel [13] conducted a survey with 49 participants, which showed that developers feel a broad range of affects while programming—all the affects of the measurement instrument's spectrum. Positive affects are perceived to be those enhancing their productivity. *Frustration* is the negative affect more often felt, as well as the one perceived as deteriorating productivity.

We also conducted two studies to understand the connection between affects and performance of software developers. The first [9] was a correlational study of real-time affects and the self-assessed productivity of eight software developers while they were programming. Their affects and their productivity were measured in intervals of 10 minutes. We found evidence for a positive correlation between affects of developers (in terms of valence and arousal) associated to a programming task and their self-assessed productivity.

In the second study [8], we recruited 42 Computer Science students to investigate the relationship between the affects of software developers and their creative and analytical performance. The participants performed two tasks chosen from psychology research, one related to creative performance, the other to analytic performance, which is resembled to algorithm design and execution. The participants' pre-existing affects were measured before each task. The study provided empirical support for the claim that happy developers are indeed better problem solvers in terms of their analytical abilities. It further raised the need for studying human factors in software engineering by employing a multidisciplinary viewpoint.

## Listening to the Voice of the Practitioners

Our second study [8], published in an open access venue, has been well received by practitioners on the social media, the news outlets, and the general Web. Four weeks after the publication, the article reached 5000 views. It gained immediate attention from different news outlets all over the world, including ITworld and the Daily Mail[2]. The attention from the social networks (Twitter, Facebook, Google+, LinkedIn) was overwhelming. The article appeared on the most important outlets for practitioners, including Slashdot, Reddit, CSDN, and Soylent News. The news articles reporting our paper, on the other hand, reported more than 1000 social networking shares. The "going viral" of the article shows that practitioners are interested and keen to discuss the linkage between happy developers and performance.

304 comments were posted on the previously mentioned news and social media channels within two months of publication. They are the Voice of Practitioners worth listening to, which we did. We analyzed the 210 comments made in English language. Many comments stated "*this is obvious.*" However, quite a few in the meantime stressed the importance to test the obvious in science, "*Before you test something you don't know, you're guessing. […] Research like this is important!*" Several comments show the practitioners understood why human aspects should have a major role in software engineering studies: "[Happiness] *has a bigger effect on software developers' productivity that it would in non-creative jobs […] When a software developer is demoralized, you can get some truly awful code.*" That is obvious. Or is it not? Several practitioners correctly understood that "*the study isn't about whether music and a mini-bar make better programmers, it's about whether being happy makes one a better programmer.*"

There were also a high number of developers reporting how they were feeling bad in their companies, and were "*contemplating quitting over mistreatment.*" Others even quitted their jobs because they felt unhappy there, and unproductive: "*At that point the motivating factor was making the project a success for another checkmark in my resume only... so I could leverage that for a position at a new company that would treat their developers better.*" While the majority of the source of unhappiness at work was related to the sources of interruption, the most surprising one was the dissatisfaction with open spaces and forced communication with other people. "*I got a lot more done in my quiet 6x6 cube than in a 16x32 open plan shared with eight other guys.*" The critical article by Skowronski [14] supports these comments by criticizing overly strict application of Agile without actually focusing on needs of single individuals. On the other hand, practitioners appear to be willing to share what makes them feel happy at work. Among the top cited are a quiet environment, limitation of multitasking, the possibility of performing frequent short breaks, and a comfort zone with coffee near the office.

---

[2] The links are collected here: https://peerj.com/articles/289/#links

According to practitioners, managers need to understand the uniqueness of people. *"For some people, music and a mini-bar make them happy, for others, it is silence and being free to develop. It comes down to 'When you get what makes you happy, you will work better.'"* In addition, *"The problem is that even if HR or PHB understand this, they may try to apply a one size fits all methodology to engender happiness."* An experienced software engineer, referring to an experience with a well-known semiconductor company, left what we think is an inspiring comment.

*"I've never worked so hard, put in more hours, got more stuff done, cranked out more code, etc, as I have in my [Company Name] time. Why? In meetings my ideas were listened to. I had a ton of freedom in my job to Get Things Done. I was recognized for Stuff I Got Done. I was not bogged down in daily staff meetings, weekly department meetings, etc. I had input on who to hire for my team. Most of all, I Had A Door I Could Close (but never did). Treat your employees like intelligent people, give them the tools they need, get out of the way, and they will not only be happy, but productive as [censored]."*

Giving simple managerial advice is difficult since positive psychological effects such as happiness are not easily introduced in a workplace. Science has shown that practitioners feel a broad range of affects while developing software, and that their moods and emotions can change dramatically in a short time period. However, evidence hints clearly that positive affects pay off in terms of performance. What is yet to be known is how to foster positive affects of software developers. Science whispers only very few concrete advices like allowing developers to listen to music while developing. Something more concrete is emerging from the Voice of Practitioners, such as a quiet work environment, limitation of multitasking, and the possibility of having short breaks often. However, we want to remind the readers that single gimmicks are not guaranteed to work. We feel that the solution is actually voiced out by the practitioners themselves. The secret is to respect, treat and value each voice as unique individuals. While this is more easily said than done, the managers should know that the developers are very willing to share their ideas, concerns and suggestions.

Daniel Graziotin is a PhD student in Computer Science at the Free University of Bozen-Bolzano. His research focuses on human aspects in software development, empirical software engineering with psychological measurements, Web engineering, and Open Science. His ORCID is 0000-0002-9107-7681. Contact him at daniel.graziotin@unibz.it

Dr. Xiaofeng Wang is a researcher at the Free University of Bozen-Bolzano. Her research areas include software development process and methods, agile development, and complex adaptive systems theory. Contact her at xiaofeng.wang@unibz.it.

Prof. Pekka Abrahamsson is a full professor of computer science at Free University of Bozen-Bolzano. His research interests are centered on empirical software engineering, agile development, startups and cloud computing. Contact him at pekka.abrahamsson@unibz.it.